\documentclass[twoside]{article}
\usepackage{graphicx}
\usepackage{fancyhdr}
\usepackage{multicol}
\usepackage{styl-ap}
\usepackage[T1]{fontenc}

\begin{document}
\title{The role of magnetic field for quiescence-outburst models in CVs}
\author{S. De Bianchi\work{1,2}, V. F. Braga\work{3}, S. Gaudenzi\work{1}}
\workplace{Dept. of Physics, Universit\`a degli Studi di Roma "La Sapienza", Piazzale A. Moro 5, 00185 Rome, Italy
\next
{ \'{E}cole Normale Sup\'{e}rieure, CNRS-UMR 8547, Rue d'Ulm 45, 75005 Paris, France}
\next
Dept. of Physics, Universit\`a Tor Vergata, Via della Ricerca Scientifica 1, 00133 Rome, Italy}
\mainauthor{silvia.debianchi@uniroma1.it}
\maketitle

\begin{abstract}%
In this paper we present the elementary assumptions of our research on the role of the magnetic field in modelling the quiescence-outbursts cycle in Cataclysmic Variables (CVs). The behaviour of the magnetic field is crucial not only to integrate the disk instability model (Osaki 1974), but also to determine the cause and effect nexus among parameters affecting the behavior of complex systems. On the ground of our interpretation of the results emerging from the literature, we suggest that in models describing DNe outbursts, such as the disk instability model, the secondary instability model (Bath 1973) and the thermonuclear runaway model (Mitrofanov 1978), the role of the magnetic field is at least twofold. On the one hand, it activates a specific dynamic pathway for the accreting matter by channelling it. On the other hand, it could be indirectly responsible for switching a particular outburst modality. In order to represent these two roles of the magnetic field, we need to integrate the disk instability model by looking at the global behaviour of the system under analysis. Stochastic resonance in dynamo models, we believe, is a suitable candidate for accomplishing this task. We shall present the MHD model including this mechanism elsewhere.
\end{abstract}


\begin{multicols}{2}
\section{Introduction}
In the last decades, methods and models used to describe DNe outburst cycles achieved several results (Bath \& van Paradijs 1983, Cannizzo \& Mattei 1998, Cannizzo 2012). However, there is no unique model that predicts these cycles (Smith 2007), as well as their underlying mechanism. The limit-cycle model for dwarf nova outbursts is generally accepted, but there are still uncertainties about how to explain all the details, especially for the SU UMa stars (Smith 2007). Even if we know that the secondary stars in DNe are magnetically active, yet it is unknown how they maintain a dynamo in the presence of tidal forces or whether there is differential rotation (Smith 2007). Crawford et al. (2008), for instance, when presenting their results of the detection of the first observed outburst of DW Cnc, asked the question of what caused the outburst observed. DW Cnc experienced a magnitude change of $\sim$4 mag showing a behavior similar to those systems whose outbursts are due to disk instability. However, with respect to $P_{spin}$  and $P_{orb}$, DW Cnc shares characteristics of those systems whose outbursts are due to a mass transfer event. Therefore, a decision for which mechanism was responsible for the first detected outburst could not be made. Precisely in order to overcome this kind of problems of undecidability a deeper analysis of our models' ability to explain DNe outbursts is needed.

\section{Disk instability model and secondary instability model: some open questions}
Each of our best available models, such as the disk instability model (DIM), the secondary instability model (SIM) and thermonuclear runaway (TNR) model, taken separately, shows weaknesses in predicting the outburst-quiescence cycles of many systems, such as V513 Cas for outbursts during standstills (Hameury \& Lasota 2014). DIM is generally accepted as reproducing an explanation of DNe outbursts and invokes an intrinsic modulation of the accretion rate in the disk. However, as seen in the case of DW Cnc, the model does not satisfactorily account for all systems and suffers of some structural problems. The first one has been inherited by the disk model of Shakura \& Sunyaev (1973) which was based on a constant value of ${\alpha}$, a parameter that stores all unknown information of the complex friction processes. A constant ${\alpha}$ does not account for a reliable outburst amplitude (Smak, 1984) and the parameter appeared to be an ad hoc solution for strictly regular outbursts (Bath 2004). Studies concerning its variation or a turbulent ${\alpha}$ parameter aimed to the appropriate simulation of the disk instability by applying suitable corrections (Latter \& Papaloizou 2012; Penna et al. 2013; Potter \& Balbus 2014). A second problem consists in the fact that DIM does not appropriately model the observed fluctuations, due to the assumption of a constant value for mass transfer rate \.{M}. Third, DIM shows weaknesses in predicting global changes in the disk structure, such as whether thin disk accretion can make a transition to Advection-Dominated Flow (Narayan \& Yi 1994, 1995a, 1995b) or how an accretion disk creates and powers jets at its center (King 2012). As an alternative to DIM, SIM assumes a mechanism triggering the outbursts based on mass transfer modulation from the secondary star (Bath et al. 1974). By combining dynamical instabilities of the secondary, time evolution of the accretion disk together with thermonuclear burning due to accreted material, SIM produces reliable results: outbursts during standstills in Z Cam systems can be explained only by appealing to instabilities in the flow from the secondary (Hameury \& Lasota 2014). It seems then that the SIM is decisive in predicting CVs like Z Cam variables and, possibly, SU UMa variables. Nevertheless, its application has not been always straightforward. Osaki (1985), for instance, explained the superoutbursts of SU UMa variables by using SIM, but later denied this possibility in favor of a thermal-tidal-instability (TTI) model (Osaki 1989). Finally, Smak (1996) suggested a hybrid TTI-SIM model, which is still debated today (Osaki \& Kato 2013).

\section{Magnetic field and outburst modality}
In order to enrich the theoretical scenarios in dealing with DNe outburst-quiescence cycle, the (TNR) model shows interesting implications, even if the improvement of the TNR model launched by Shara (1982) never brought to a decisive point. TNR introduces a time parameter describing the recurrence of outbursts accounting for both Classical Novae (CNe) and DNe outbursts and their differences (Shara 1982) and also accounts for DNOs and X-ray flux variations concerning some representative systems like U Gem, SS Cyg, EX Hya, Z Cam, CY Cnc, AH Her, CN Ori, KT Per. More importantly, TNR can be associated to one of the processes that causes Nitrogen to Carbon enhancement observed on the surface of the WD in VW Hyi and U Gem (Sion 2014). In TNR models, outbursts are the product of thermonuclear burning onto the WD surface (Mitrofanov 1978; 1980). In the presence of a magnetic field of the order of $3\cdot 10^6-3\cdot 10^7$ G on the surface of the degenerate component, a local accumulation of hydrogen could generate, through thermal instability, a thermonuclear burning responsible of the outburst. For this process to occur, an intense magnetic field $10^6-10^8$ G is required; primaries with such fields do exist in Intermediate polars (IPs) and Polars that owe their properties to the very primary magnetic field (Patterson 1994). Therefore, the magnetic field strength on the WD surface allows us to distinguish novae, dwarf novae and novalike stars. In exploring models that go beyond the use of TNR alone, Livio (1983) emphasized the importance of the magnetic fields in CVs (Livio \& Verbunt 1988; Meyer-Hofmeister et al. 1996). As Livio (1983) shows, nova explosions can be inhibited if the magnetic field strength is over a certain limit, explaining in this way why active novae are absent among Polars while novae can be found among the IPs.

\begin{myfigure}
\centerline{\resizebox{70mm}{!}{\includegraphics{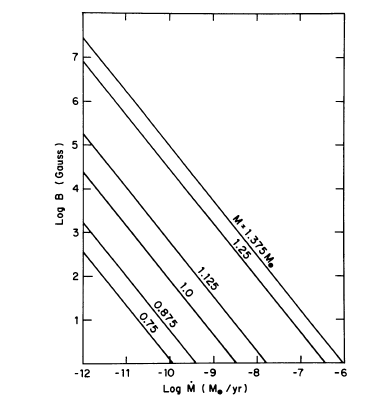}}}
\caption{Upper limits on the magnetic field strength for which nova outbursts can occur (provided that matter is confined to polar caps) as a function of the accretion rate, for various white dwarf masses. Reproduced from Livio, M., 1983, Astronomy and Astrophysics, 121, L7, with permission from Astronomy and Astrophysics, \copyright ESO.}
\label{Livio 1983-fig2}
\end{myfigure}

Livio's approach allows one to determine upper limits on the magnetic field strength for which nova outbursts can occur when polar caps are present. These upper limits can be represented as a function of the accretion rate and for different WD masses (see Figure 1). Even if a magnetic field more intense than a certain critical value could inhibit nova explosions (Livio 1983), a magnetic field is strictly correlated, on an intensity-dependent scale, to different modes of accretion and thermonuclear runaways. These modes include the channeling of the accreted matter into magnetic polar caps, the alteration of radiative and conductive opacities, interference with the development of convection that likely is also present in non magnetic CVs. Furthermore, the magnetic field of the WD itself may affect the nova outbursts, e.g. by enhancing mass loss in the equatorial plane (Livio et al. 1988; Prialnik \& Livio 1995). This reveals how the magnetic field activates different and specific dynamic pathways. In other words, to focus on the role of a variable magnetic field and its effects can be helpful for identifying common features of both magnetic and non-magnetic CVs, as well as for obtaining crucial information on the magnetic field responsible for switching a particular outburst modality. Indeed, according to Livio (1983), the pressure at the base of the accreted matter is the physical parameter that determines the outcomes of a TNR:

\begin{equation}
P_b=\frac{GM_{WD}}{{R^2}_{WD}}\frac{\Delta m_{acc}}{A_{cap}}
\end{equation}

When $P_b$ exceeds a critical value of some $10^{19} dyne/cm^2$, ignition occurs. It should be noted that, for higher CNO abundances, the critical pressure for the occurrence of an outburst is lower. This fact leads us to the investigation of CNO abundances and of the presence of material accumulating during quiescence, in order to speculate on the triggering of an outburst. It means that a function of ``prediction'' and not of mere ``description'' can be added to the model. 

\begin{myfigure}
\centerline{\resizebox{70mm}{!}{\includegraphics{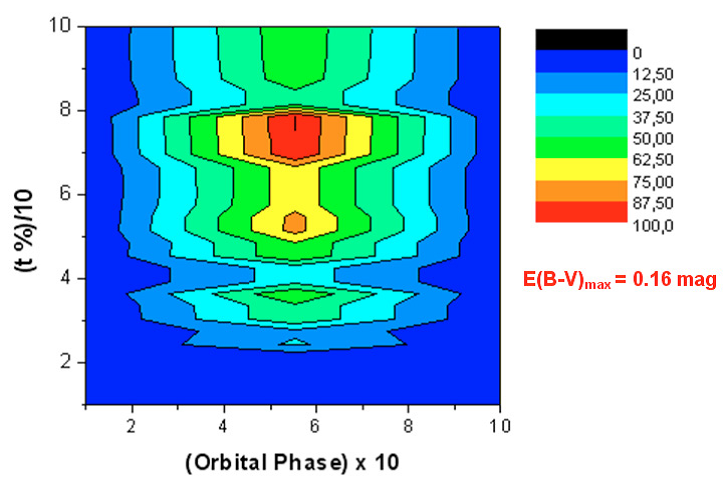}}}
\caption{ E(B-V) varies with both the orbital phase and the percentile quiescence time ({t}\%). 2175 \textup{\AA} absorption bump is believed to be caused by $C_{60}$. E(B-V) takes the highest value at the ``heart'' of the quiescence. $C_{60}$ molecules could be reasonably supposed to be affecting the modulation of time elapsing between the outbursts. Reproduced from Gaudenzi et al. (2011), Astronomy and Astrophysics 525, A147 with permission from Astronomy and Astrophysics, \copyright ESO.}
\label{Gaudenzi et al. 2011-fig3}
\end{myfigure}

More importantly, the analysis of UV spectra revealed the presence of fullerenes as an intrinsic source of reddening in SS Cyg (Gaudenzi et al. 2011, see Figure 2). Density gradients in the disk may influence the accumulation of molecules in specific sites of the disk itself where they are accreted and/or ejected during the quiescence-outburst cycle. Along with accretion/ejection mechanisms, a major role in determining the structure and the stability of the disk might be played by the solid-gaseous phase transitions of fullerenes (Gaudenzi et al. 2012) at 5855 K (Hussien et al. 2008). On the ground of the hypothesis that the quiescence-outburst phase transition is the effect of a stochastic resonance mechanism, the presence of material accumulated in the disk could trigger wave amplification within the disk even during quiescence. This would introduce in our modelling of the disk instability an amplification of stochastic resonance induced by turbulent fluctuations (Benzi \& Pinton 2011). 

\section{Magnetic field and accretion behaviour}
Seminal numerical studies of the influences of a magnetic field on the flow structure were performed in the early 1990s either in the frame of simplified models (King 1993; Wynn \& King 1995; Wynn et al. 1997; King \& Wynn 1999; Norton et al. 2004; Ikhsanov et al. 2004; Norton et al. 2008) or in a limited region of the stellar magnetosphere (Koldoba et al. 2002; Romanova et al. 2003, 2004). In the last few years, there have been attempts to develop a comprehensive 3D numerical model to calculate the flow structure in close binaries (Zhilkin \& Bisikalo 2009, 2010; Bisikalo \& Zhilkin 2012). According to our interpretation of the results of 3D MHD simulation of Bisikalo \& Zhilkin (2012), it emerges that the value of the magnetic induction on the surface of the accreting star activates the dynamic pathway and, in doing so, distinguishes two different modalities of accretion. In the numerical model,  Bisikalo \& Zhilkin  (2012) take into account radiative heating and cooling,as wells as diffusion of the magnetic field due to dissipation of currents in turbulent vortexes, magnetic buoyancy, and wave MHD turbulence. The interesting result consists in that if the magnetic field induction grows, the cross-section of the stream decreases and the accretion rate decreases as well, thereby influencing density and pressure, and thus affecting the possibility of triggering an outburst. We interpret the non-monotonic variation of the magnetic field behaviour and its consequences reported in (Bisikalo \& Zhilkin 2012) as stochastic oscillations of B. In particular, magnetic field values around $10^6 G$ could originate phase transitions between polar and IP. Based on our previous work (Gaudenzi et al. 2012), we suggest that the dynamic pathway of accretion can account for instabilities of many CVs and discloses the possibility of acquiring an ability to predict their behaviour as transition objects. The physics behind this process concerns the amplification of stochastic resonance induced by turbulent fluctuations, i.e. the amplitude of the external periodic perturbation needed for stochastic resonance to occur is much smaller than the one estimated by the equilibrium probability distribution of the unperturbed system (Benzi \& Pinton 2011). 

\section{Discussion and conclusion}
In order to enrich the theoretical scenario we stressed the relevance of the accretion process that may change the surface magnetic field of an accreting WD significantly (Cumming 2002), and, in agreement with (Livio 1983), that the magnetic field of the WD itself may affect the nova outburst. We also remarked that the accumulation of some type of molecules, such as $C_{60}$ molecules within the disk is important to understand the behavior of DNe in quiescence and that the CNO abundances also depend on the presence of a magnetic field (even of a weak one). If we want to predict the behavior of a complex system, such as DNe outbursts cycles, neither DIM nor any other model taken alone is sufficient. In fact, DIM explains why thermal instability leads to outbursts, but it does not provide exact information at the level of the global system in the quiescent phase, e.g. whether there are transition from a state to another, and how to determine time-dependent ${\alpha}$ variations. In order to obtain more information about the DNe outburst-quiescence cycle, we shall work on a model that is able to predict future states of these systems, as well as to describe the present state, without appealing to ad hoc assumptions. As previously stated, local, small-scale turbulences, may likely represent the physical outcome of stochastic oscillations of the magnetic field. Among the other mechanisms, such as a strong differential rotation and vertical density gradients (Pudritz 1981b), turbulences can be responsible for the generation of a large-scale magnetic field (Pudritz 1981a). Therefore, the development of mean field dynamo theory is crucial in order to overcome some difficulties related to current models in accounting for the role of the magnetic field generated in a turbulent medium. This theory, if appropriately modified, could allow us to calculate the overall structure of the global field without any detailed knowledge of the small-scale turbulence. Even if this echoes earlier works by Shakura \& Sunyaev (1973) there is a characteristic of the mean field dynamo theory that makes it different from earlier theories. It builds up a large-scale magnetic field trying to achieve a global redistribution of the angular momentum, whereas other models assume a small-scale field and a local viscosity. Therefore, in order to gain knowledge on the underlying mechanism producing outbursts, such as the one presented in DW Cnc, we will 1) simulate the interaction among thermonuclear runaway and magnetic field leading to a specific modification of the ${\alpha}$ viscosity parameter to be integrated in the model (switching the outburst modality); 2) model dynamic pathways of accretion due to the magnetic field behaviour by means of a MHD model that takes into account stochastic resonance in a mean field dynamo model. We shall present and discuss this simulation elsewhere, but we presented here the theoretical assumptions that inspired our study.

\end{multicols}

\begin{multicols}{2}

\bigskip
\bigskip
\noindent {\bf}

\bigskip
\noindent {\bf } 

\bigskip
\noindent {\bf } 

\bigskip
\noindent {\bf } 

\bigskip
\noindent {\bf }

\end{multicols}
\end{document}